\documentclass[prl,twocolumn,showpacs,preprintnumbers,amsmath,amssymb,superscriptaddress]{revtex4}
\usepackage{latexsym}
\usepackage{graphicx}% Include figure files
\usepackage{pstricks,pst-node,pst-tree}
\usepackage{dcolumn}% Align table columns on decimal point
\usepackage{bm}% bold math
\usepackage{xcolor}
\usepackage{float}
\usepackage{epsfig}
\usepackage{amsmath}

\usepackage{sidecap}

\input amssym.def
\input amssym.tex

\newcommand{\bea}{\begin{eqnarray}}
\newcommand{\eea}{\end{eqnarray}}

\newcommand{\nn}{\nonumber}

\begin{document}
\title{An Exactly Solvable Discrete Stochastic Process with Correlated Properties}
\author{Jongwook Kim}
\email{dr.jongwookkim@gmail.com}
\affiliation{Asia Pacific Center for Theoretical Physics, Pohang, Korea}
\author{Junghyo Jo}
\affiliation{Asia Pacific Center for Theoretical Physics, Pohang, Korea}
\affiliation{Department of Physics, POSTECH, Pohang, Korea}

%}\footnote{Electronic correspondence:~\bf{jwkim@apctp.org}}

\begin{abstract}
We propose a correlated stochastic process of which 
the novel non-Gaussian probability mass function is constructed 
by exactly solving moment generating function. The calculation of cumulants and auto-correlation shows that
the process is convergent and scale invariant in the large but finite number limit. We demonstrate that
the model consistently explains both the distribution and the correlation of discrete financial time-series data, and predicts the data distribution with high precision in the small number regime. 
\end{abstract}


\maketitle

Non-normal distributions and time-series clustering are prevalent in physical and biological phenomena, such as in crystal growth, polymer transportation/distribution\cite{Sornette,deGennes}, and brain electrical activity\cite{LehnertzEglger}. In addition, non-normal distributions with high kurtosis (heavy-tail or leptokurtosis) and volatility clustering are frequently observed in the social science, such as in financial time-series data\cite{Kleinert,Rachev} and social networks\cite{Barabasi}\,. 
Leptokurtic distributions have been modeled by a family of stable distributions, constructed from the composite of independent and nonidentical Brownian and Poisson-jump processes.
Clustering in time-series events is caused by the correlations between these events, for which 
various continuous and discrete models have been proposed to explain their phenomena. 
Auto-regressive moving average models\cite{Rachev} are popular continuous Gaussian stochastic models with correlated properties. 
Fractional Brownian motion\cite{Mandelbrot} is a continuous stochastic process, 
whose model is defined by the exponent of the power-law auto-correlation function of two white noises at different times. 
There are also various discrete models of correlated random walks, collectively referred to as urn models. 
Classical urn schemes are Polya's model\cite{PolyaUrn}, which is implied by the $\beta$-distribution, 
and Friedman's model\cite{FriedmanUrn}, which is a kind of dual to the Polya's model. 
However, not much attention has been given to urn models in the analysis of time-series events, compared to the use of various continuous volatility models. 
Due to the inflexibility and limitations of current stochastic modelings, 
there is no unified prescription for the analysis of all kinds of data, and only few discrete processes are solved analytically. In such circumstances, it is worth developing a new discrete stochastic scheme constructed from a discrete  micro process to explain the volatility of time-series data using correlations.

\begin{figure}
\begin{centering}
\includegraphics[width=.35\textwidth]{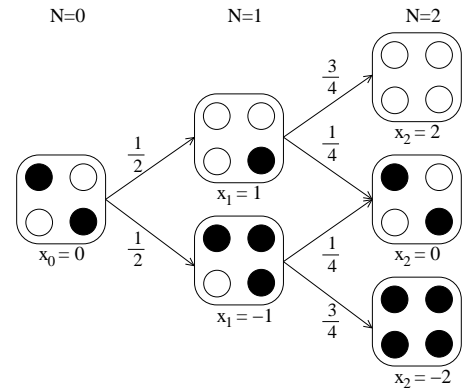}
\caption{Ensemble paths of two iterations($N=2$), when $\kappa={1/2}$. A hollow circle represents half of the total initial number of white balls and a solid circle represents the same number of black balls. For example, the drawing of a white ball (upward right direction) increases the stochastic position by $+1$ and leads to the replacement of half of the black balls into white balls. One then proceeds to the next iteration. 
}
\label{1}
\end{centering}
\end{figure}

In this Letter, we propose a binomial process of Markovian correlation. It is an urn model of ball replacement with a finite number of balls $\mathcal{N}$ and iterations $N$, which starts to converge at a large but finite number of iterations $N$. The process begins with the same $\mathcal{N}$ number of black and  white balls in the urn, assuming unskewedness. 
In Ehrenfest's urn scheme, one changes the color of a white ball to black in the case of drawing a black ball. After one round of drawing a black ball, one has $\mathcal{N}+1$ black balls and $\mathcal{N}-1$ white balls in the urn, and the probability of drawing a black ball in the second round turns to be biased as $1/2+1/2\mathcal{N}$. However, the process soon becomes ill-defined for $N$ iterations larger than $\mathcal{N}$. Therefore, Ehrenfest's urn process is well-defined only for an infinite number of balls and iterations, solved by taking $N\rightarrow \infty$ limit, and convergent to the Ornstein-Uhlenbeck distribution in the continuum limit\cite{PolyaUrn}. We generalize this process by adjusting the amount of color interchange throughout the whole process to make the process well-defined without taking $N\rightarrow \infty$ limit. Given the number of total iterations $N$, the ratio of interchange is adjusted to $2\kappa/N$ of the initial number of white/black balls $\mathcal{N}$, where $\kappa$ is the correlation parameter ranging from $-1/2$ to $1/2$. Thus, the probability of drawing a black ball after the aforementioned first round is generalized as $1/2+(\mathcal{N}\cdot 2\kappa/N)/2\mathcal{N}$. The rescaled correlation parameter is denoted as $\epsilon(N) = {\kappa / N}$ for brevity. The stochastic displacement $\delta x_n$ at the time $n$ is either $+1$ or $-1$ depending on the outcome of white or black respectively, and the stochastic position $x_n$ at the $n$th iteration is defined by the difference between the historical outcomes of the white balls and that of the black balls (Fig. \ref{1}).

Within a few hundred iterations, this process describes the conspicuous non-Gaussian properties of finite $N$ statistics. Herein, we demonstrate the utility of our model in this regime by testing it on high-frequency financial time-series data, where the large volatility and positive auto-correlation are observed as its non-Gaussian entities. As $N$ grows larger, the process becomes scale invariant so that the cumulants start to converge. In the continuum limit, the proposed process ultimately converges to the Ornstein-Uhlenbeck distribution.

The recursion for the probability mass function(PMF) $P_{x,n}$ 
with the modified binomial transition probability is given as
\bea
\!\!  P_{x,n\!+\!1}\!=\!P_{x\!+\!1,n} [{1\over 2}\!-\! {(x\! +\! 1) }{\epsilon}]  \!+\!	P_{x\!-\!1,n} [{1\over 2} \!+\! {(x\! -\! 1 )}{\epsilon}]\,, \label{recurrence}
\eea
%\bea
%P(i,n+1)&=& \Big[{1\over 2}+{(i-1) }\epsilon \Big] P(i-1,n) \nonumber \\
%&&+ \Big[{1\over 2} -{(i+1)}{\epsilon }\Big] P(i+1,n), \label{recurrence}
%\eea
where $n=\{0,1,2,\cdots,N\}$\, and each event is indexed as $n$ for the entire $N$ events. 
We define the stochastic location $x_n$ at the time $n$ 
by the accumulation of the stochastic displacements $\delta x_i$ of values $\pm 1$ from the origin, {\it i.e.} 
$x_n=\sum^n_{i=1} \delta x_i$\,, and $x_0=0$. The location $x_n$ realized at the time after $n$ runs from $-n$ to $n$ in steps of $2$, {\it i.e.} $x_n \!\in\! \{ -n\,, -n\!+\!2\,, -n\!+\!4 \,, \cdots\,, n\!-\!2\,, n\}$, and the process starts with $P_{0,0} = 1$. We omit the subscript $n$ on $x_n$, when the meaning is obvious. 
The process is the first order Markovian 
%\footnote{ The stochastic location $X_n$ is the first order Markovian variable, since
%$
%P(X_n = x_n | X_{n-1}=x_{n-1}\,, X_{n-2}=x_{n-2}\,, \cdots X_{1}=x_{1}) 
%\!=\! P(X_n =  x_n | X_{n-1}=x_{n-1})\,,
%$ where $x_i$ is the value of the historical realization. However, $\delta X_n$ is the non-Markovian stochastic variable, since it depends on all $\delta X_i$, where $i=1,2,\cdots n$\,.
%For example, let us denote the conditional probability of getting $+1$ at the time $3$ as $P(\delta X_3 \!=\! 1 | \cdots)$, and then
%$
%P(\delta X_3 \!=\! 1 | \delta X_2\!=\! 1)\!=\!{1\over 2} \!+\! (x + 1)\epsilon_N   $\,, where $ x\in \{1, -1\}$$\,,$ and $
%P(\delta X_3 \!=\! 1 | \delta X_2\!=\! 1,\delta X_1 \!=\! 1)\!=\!{1\over 2} \!+\! 2 \epsilon_N$. In such a way, the Polya's urn is sometimes referred to be a non-Markovian process\cite{nonMarkovUrn}.}\, 
and becomes the Ornstein-Uhlenbeck equation with the density function $p(x,t)$ 
in the continuous limits of position and time: 
$\partial_t p = -2\epsilon \partial_x(xp)+{1/2}\partial_x^2 p$.

The moment generating function is introduced as
$Z_n(q)= {\sum}_{x} q^x P_{x,n}\,,$ where $Z_n(0)={\sum}_{x} P_{x,n} = 1$\,.
The moment generating function at time $0$ is defined as $Z_0(q)=1 \label{BCs}$, and then the recurrence in Eq.~(\ref{recurrence}) is recast into a differential equation for $q$, shown as 
\bea
Z_{n+1}(q) = {1\over 2}(q+q^{-1}) Z_n(q)
		+ {\epsilon }(q^2-1) \partial_q Z_n(q) \,. \label{Zequationlinear_x}
\eea
The substitution of $Z_n(q) = \epsilon^n (q - q^{-1})^{-{1/ 2\epsilon}}Y_n(q)$, leads to the simple equation
\bea
Y_{n+1}(q) = 
			(q^2-1) \partial_q Y_n(q) \,
\eea
with $Y_0(q) = (q -q^{-1})^{1/ 2\epsilon}$\,.
A variable change of $q=-\tanh(r)$\, gives the simpler equation, $Y_{n+1}(r) = \partial_r Y_n(r)$
with $Y_0(r) = [{\sinh(2r)/ 2}]^{-{1/ 2\epsilon}}$.
Then, the partition function $Z_n(r)$ in the $r$ coordinate is
\bea
Z_n(r) = \epsilon^n \Big[ \frac{\sinh(2r)}{2} \Big]^{\frac{1}{2\epsilon}} \partial_r^n \Big[ \frac{\sinh(2r)}{2} \Big]^{-\frac{1}{2\epsilon}}.
\eea

\begin{figure}
\begin{centering}
\includegraphics[width=.48\textwidth]{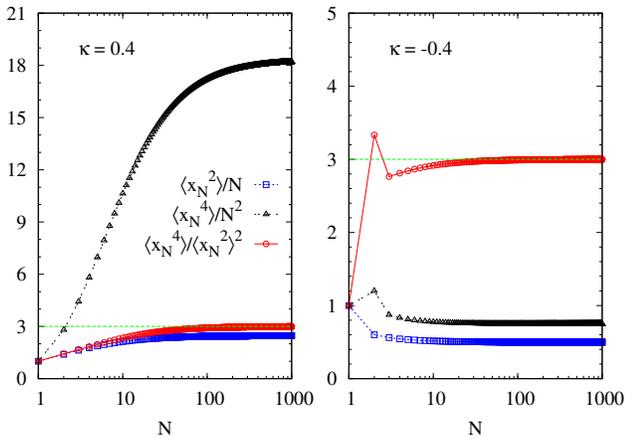}
\caption{(Color online) Variance (square, blue), the fourth moment (triangle, black) and kurtosis (circle, red) 
as a function of $N$ with $\kappa=0.4$ (left) and $\kappa=-0.4$ (right).
A guide line (y=3, green) is drawn as a reference.
}
\label{fig2}
\end{centering}
\end{figure}

\begin{widetext}
Using the identity $\partial_r^n \sinh^\alpha(y)=\alpha {{-\alpha+n} \choose n} \sum_{k=1}^n (-1)^k (\alpha-k)^{-1} \sinh^{\alpha-k} (y) \partial_y^n \sinh^k(y) $, the closed form of the generating function is obtained as\,
%\begin{widetext}
\bea
\!\!\!&&\!\!\!\!\!\!\!\!\!\!\!\!\!\!\! Z_n(q) = 
\sum_{k=1}^n \sum_{i=0}^k
 {{{1/2\epsilon}+n}\choose{n}}
 {{n}\choose{k}}
 {{k}\choose{i}}
 \frac{(2\epsilon)^{n-1}  (2i -k)^n}{({1/2\epsilon}+k)}
 \Big[-\frac{(1+q)^2}{4q} \Big]^k
 \Big[-\frac{(1-q)^2}{(1+q)^2} \Big]^i\,,
 \label{Zq}
\eea
\end{widetext}
where the binomial function is defined as the product
${{{1/ 2\epsilon}+n}\choose{n} } = {\prod_{i=0}^{n-1}({1/ 2\epsilon}+n-i)/ n!}$. 
The Gaussian model is recovered with the absence of correlation, and therefore  Eq.~(\ref{Zq}) becomes the familiar moment generating function of Bernoulli process; $
\displaystyle \lim_{\epsilon \to 0} Z_n(q) = {1/ 2^n} (q+q^{-1})^n \,.
$
%
%\begin{figure}
%\begin{centering}
%\includegraphics[width=.40\textwidth]{fig2.eps}
%\caption{Probability distribution functions $P_{N|x}$ with 
%$N=40$ (left) and $N=100$ (right). The two values of $\epsilon_N$ are chosen to have the same $\kappa=\pm 0.4$.  The Gaussian distribution where $\kappa=0$ is also shown for comparison.  
%}
%\label{fig2}
%\end{centering}
%\end{figure}
%
%[[[[ TAKE OUT !! The coupling $\epsilon_N$ is fixed at all times $n$ but rescaled by the totally elapsed time $N$ as ${\kappa / N}$. We prove that the stochastic process is scale invariant under such rescaling, by showing the convergence of cumulants. The probability distributions for the fixed $\kappa = \pm 0.4$ with different time scales $N=40$ and $N=100$ are illustrated in Fig. \ref{fig2}\,, indicating the scale invariance.  Notice that the positive correlation $\kappa$ predicts faster diffusion speed then the binomial process, and vice versa for the negative correlation. ]]]]]
%
% 
%  
Cumulants are calculated from the derivatives of the generating function Eq.~(\ref{Zq})\,. 
As expected, the average and the skewness are both zero. 
The variance and the fourth moment at the final time $N$, in the limit of small $\kappa$, are calculated to be
%\bea
%\!\!\!\!\!\!\mathbb{E}[ X_{N}^2 ]
%\!\!\!\!&=&\!\!\!\! N\!\! + \!\!{4N\!(N\!\!-\!\!1)\over 2!} {\kappa\over N} \!\!+\!\! {{16}N\!(N\!\!-\!\!1)\!(N\!\!-\!\!2)\over 3!}\!(\!{\kappa\over N}\!)^2 \!\!\!+\!\mathcal{\!O}(\!\kappa^3\!),\!\!\label{var}
%\\
%\!\!\!\!\!\!\mathbb{E}[ X_{N}^4 ]
%\!\!\!\!&=&\!\!\!\! N(3N-2)\!\! + 8{N(N\!\!-\!\!1)\over 2!}(3N\!\!-\!\!4){\kappa\over N}  \nn\\
%\!\!\!& &\!\!\! \!\!+56\, {N(N\!\!-\!\!1)(N\!\!-\!\!2) \over 3!}(3N\!\! -\!\!{43\over7}) ({\kappa\over N})^2 \!+\!\mathcal{O}(\kappa^3)\,.   \label{kur}
%\eea
\bea
\!\!\!\!\!\!\langle x_N^2 \rangle \!\!\!\!&=&\!\!\!\!  [\partial_q^2 Z_N + \partial_q Z_N ]|_{q=1} \nn \\
\!\!\!\!&=&\!\!\!\! N \!\!+\!\! {4N\!(N\!\!-\!\!1)\over 2!}{\kappa\over N} \!\!+\!\!{{16}N\!(N\!\!-\!\!1)\!(N\!\!-\!\!2)\over 3!}\!(\!{\kappa\over N}\!)^2 \!\!+\mathcal{\!O}(\!\kappa^3\!),\!\!\label{var}
\\
\!\!\!\!\!\!\langle x_{N}^4 \rangle \!\!\!\!&=&\!\!\!\! [\partial_q^4 Z_N + 6\partial_q^3 Z_N +7\partial_q^2 Z_N +  \partial_q Z_N ]|_{q=1}\nn \\
\!\!\!\!&=&\!\!\!\! N(3N-2) + {8N(N\!\!-\!\!1)\over 2!}(3N\!\!-\!\!4){\kappa\over N} \nn \\
\!\!\!\!& &\!\!\!\!+{56N(N\!\!-\!\!1)(N\!\!-\!\!2) \over 3!}(3N\!\! -\!\!{43\over7}) ({\kappa\over N})^2 +\mathcal{\!O}(\!\kappa^3\!). \label{kur}
\eea
Next, we directly iterate the stochastic process in Eq.~(\ref{recurrence}) to compute the variance, fourth moment, and kurtosis of the PMF 
with positive and negative correlations with time (Fig.~\ref{fig2}).
They all grow non-linearly within several hundred iterations. The stochastic model in the small $N$ regime is used to explain positive correlations in financial time-series data, the details of which are discussed later in the text.

\begin{figure}
\begin{centering}
\includegraphics[width=.48\textwidth]{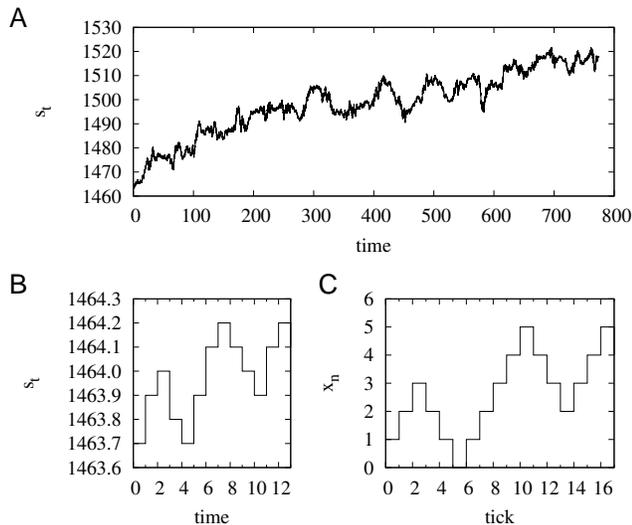}
\caption{(A) Standard $\&$ Poors futures prices of all transaction records\cite{data} $s_t$, where $t$ indicates each transaction event.
(B) The data enlarged over a short period.
(C) Tick movement decomposed from the time-series data of (B).
The minimum tick movement is set as $\pm 1$ by rescaling the minimum tick size $\pm 0.1$.
The integer-valued time-series from the raw data is denoted as $\{s_1,s_2,\cdots,s_t,\cdots\}$, 
and the time-series difference is denoted as $\{s_2-s_1, s_3-s_2,\cdots,s_{t+1}-s_t,\cdots\}$. 
We decompose the time-series of differences into a series of $\pm1$ of the length $|s_{t+1}-s_t|$,
depending on the sign of $s_{t+1}-s_t$.
For example, the time-series $\cdots 1463.7, 1463.9, 1464.0, 1463.8,\cdots$
is mapped into the tick movements $\cdots (+1, +1), +1,( -1 ,-1),\cdots$ in our analysis.
}
\label{fig3}
\end{centering}
\end{figure}

In the large $N$ regime, the $i$th cumulants scaled by $N^{i/2}$ are independent of the number of iterations $N$ 
but only dependent on $\kappa$. Thus, $\kappa$ is the characteristic parameter to determine the PMF profile and there exists a scale invariance. They are shown to converge up to the second order of $\kappa$ in Eqs.~(\ref{var}) and (\ref{kur}), 
and we resort to numerical simulations for the higher order verification (Fig.~\ref{fig2}). 
When the number of iterations is large ($10^3 \lesssim N$), the variance grows linearly with $N$
and it is rewritten as $\langle x^2_{N} \rangle \!=\! H^2  N$, 
where $ H^2\equiv 1+2\kappa + 8\kappa^2{/3} + \mathcal{O}(\kappa^3) $\,. 
In the Gaussian limit ($\kappa = 0$), the diffusion speed is $H=1$\,. 
%[[The Polya's urn model, for example, the exponent is $1$ and the amplitude $H$ depends on the correlation.]] 
For $\kappa > 0$, where the stochastic process is positively correlated, $H>1$ and the process diffuses faster than the Brownian motion, whereas for $\kappa < 0$, where the process is negatively correlated, $H<1$ and the process is less diffusive. 
The diffusion speed $H$, calculated from the Eq.~(\ref{var}) 
up to the order of $\kappa^2$ at the large $N$ regime, are $2.23$ for $\kappa=0.4$ 
and $0.63$ for $\kappa=-0.4$\,, which are consistent with the numerical results in Fig. \ref{fig3}.
In the calculation of kurtosis, nontrivial cancellations occur at each coefficient of $\kappa$ and $\kappa^2$, and consequently the quantity converges to $3$.
\bea
\lim_{N \to \infty}{\langle x_{N}^4 \rangle \over \langle x_{N}^2 \rangle^2} = 3 +\mathcal{O}(\kappa^3) \,. \label{kur_kappa}
\eea
The numerically measured kurtosis in Fig.\ref{fig3} also confirms the convergence to $3$ at all orders of $\kappa$, 
which strongly corroborates the analytic result in Eq.~(\ref{kur_kappa}). 

The auto-correlation between tick displacements 
($\delta x_i = \pm 1$) at time $n$ and time $n+l$ with a generic time lag $l=\{1\,,2\,,\cdots\}$ 
can be approximated as $\langle \delta x_{n} \delta x_{n+l} \rangle$ =  
${2 \kappa / N}+(H^2 n+l-1)({2 \kappa / N})^2+\mathcal{O}({1/ N^3})$. At the large $N$ regime, the auto-correlation is renormalized as the order of $1/N$ in the same way as the rescaling of correlation $\epsilon=\kappa/N$, which again verifies the scale invariance. In addition, it is independent of the time parameter $n$, therefore the auto-correlation of tick displacements is stationary.

\begin{figure*}[!ht]
\begin{centering}
\includegraphics[width=0.9\textwidth]{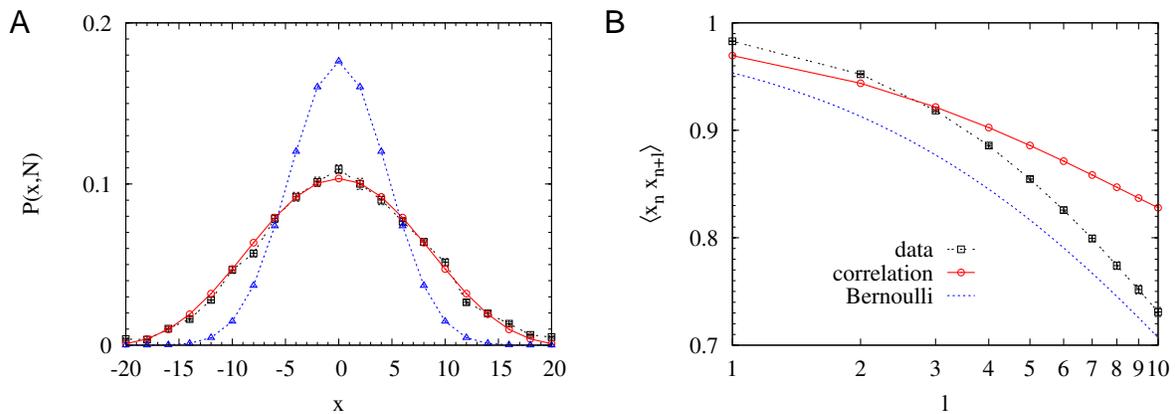}
\caption{(Color online) Frequency and auto-correlation plots of our financial data.
(A) Frequency of cumulative tick movement for $N=20$ consecutive times
and (B) auto-correlation between $n=10$ and $n+l$ consecutive tick movements
with $l=\{1,2,...,10\}$; 
financial data (square, black); Bernoulli process in the absence of correlation
between events (triangle, blue); correlation model with $\kappa=0.45$ (circle, red).
For both plots, we used $10^4$ samples of $x=x_N$ for (A) frequency and
also $10^4$ samples of $x_n x_{n+l}$ from the financial tick series.
For their uncertainty estimation, we used 100 ensembles of the given $10^4$ samples.
Note that data uncertainty (standard deviation) is smaller than the symbol size.
The auto-correlation for the Bernoulli process is $\sqrt{(n/n+l)}$.
}
\label{fig4}
\end{centering}
\end{figure*}

We test the model on high-frequency financial time-series data\cite{data} (Fig. \ref{fig3}A).
Instead of the commonly used return statistics\cite{Kleinert,Rachev}, we directly employ discrete time-series statistics of tick movements. The collection of $N$ consecutive tick movements is assumed as a statistical ensemble so that it corresponds to $x_N$(Fig. \ref{fig3}B,C). 
%and the number of data points should be at least as large as $N\cdot2^N$. 
The distribution of such ensemble data $P(x)$, where $x=x_N$ is the sum of $N$ consecutive tick displacements
$\delta x=\pm1$, has a characteristic profile with large volatility(Fig. \ref{fig4}A). Furthermore, positive correlation is observed(Fig. \ref{fig4}B).
It is possible to relate the correlation between tick movements to the non-Gaussian profile of the data, since the PMF profile of the data is clearly different from the Bernoulli process where consecutive events are independent. Therefore, we examine whether the PMF $P(x| \epsilon,M)$ of our model $M$ with the correlation parameter $\epsilon(N)=\kappa/N$ can generate the ${P}(x) \pm \delta P(x)$ of the data. The discrepancy between the model and the data is defined as
\bea
E(\epsilon)=\sum_x \frac{[{P}(x) - P(x|\epsilon,M)]^2}{2 \delta^2 {P}(x)}. 
\label{E1}
\eea
Given the mean $P(x)$ and uncertainty $\pm \delta P(x)$ of the data $D$, the model likelihood becomes $P(D|\epsilon, M) \propto \exp[-E(\epsilon)]$
due to the maximal entropy principle\cite{Sivia}.
Using a uniform prior assumption, about the parameter $\epsilon$,
{\it i.e.} $P(\epsilon|M)$=constant, the posterior probability of $\epsilon$ given data $D$
can be written as
\bea
P(\epsilon|D,M) = P(\epsilon|M) \frac{P(D|\epsilon,M)}{P(D|M)}=\frac{\exp[-E(\epsilon)]}{\int d\epsilon \exp[-E(\epsilon)] }
\eea
by the product rule in probability theory.
Then, it is straightforward to compute the expectation value 
$\langle \epsilon \rangle=\int d\epsilon \, \epsilon \, P(\epsilon | D,M)$
and its uncertainty 
$\delta \epsilon= [\int d\epsilon \, \epsilon^2 \, P(\epsilon | D,M) - \langle \epsilon \rangle^2]^{1/2}$.
We estimate the correlation parameter from the financial data using Markov Chain Monte-Carlo (MCMC) method\cite{Gregory}
with $10^5$ MC steps after equilibration:
$\epsilon(10)=0.0475\pm 0.0003$, $\epsilon(20)=0.0229 \pm 0.0001$, and $\epsilon(30)=0.0154 \pm 0.0000(5)$
for $N$=10, 20, and 30, respectively. However, the corresponding values of the correlation parameter $\kappa = \epsilon N$ are similar to each other
(0.48, 0.46, and 0.46, repeatedly) which confirms the robustness of the obtained data distribution (Fig. 4A).
%]]] 
%
%
%
%
%Given the mean and standard deviation data, $D$: $\hat{P}_N(x) \pm \delta \hat{P_N}(x)$,
%the likelihood of model is ${\cal{P}}(D|\epsilon_N, M) \propto \exp[-E(\epsilon_N)]$
%due to the maximal entropy principle (ref).
%The posterior probability,  ${\cal{P}}(\epsilon_N | D,M)={\cal{P}}(D|\epsilon_N,M) {\cal{P}}(\epsilon_N|M)/ {\cal{P}}(D|M)$,
%can be written as
%\bea
%{\cal{P}}(\epsilon_N | D,M)=\frac{\exp[-E(\epsilon_N)]}{\int d\epsilon_N \exp[-E(\epsilon_N)] }
%\eea
%with a uniform prior ${\cal{P}}(\epsilon_N|M)$=constant for the model parameter $\epsilon_N$.
%Then, we compute the expectation value $\mathbb{E}[  \epsilon_N ]$ and its uncertainty $\delta \mathbb{E}[ \epsilon_N ]$:
%\bea
%\mathbb{E}[  \epsilon_N ] &=&  \int d\epsilon_N \epsilon_N {\cal{P}}(\epsilon_N | D,M), \\
%\delta \mathbb{E}[  \epsilon_N]^2 &=& \int d\epsilon_N \epsilon^2_N {\cal{P}}(\epsilon_N | D,M) - \mathbb{E}[  \epsilon_N ]^2.
%\eea
%The model parameters for explaining the financial data are
%$\epsilon_{10}=0.0481\pm 0.0005$, $\epsilon_{20}=0.0228 \pm 0.0002$, and $\epsilon_{30}=0.0154 \pm 0.0002$.
%Here $\kappa_H = N \epsilon_N$ has consistent values as 0.48, 0.45, and 0.46 for different $N$.

%\begin{widetext}
%\end{widetext}

In the model, we also calculate the auto-correlation 
between the stochastic positions $x_n$ and $x_{n+l}$ with time lag $l$ as
\bea
{\langle x_{n} x_{n+l} \rangle \over \sqrt{\langle x_{n}^2 \rangle \langle x_{n+l}^2 \rangle}}
= \sqrt{\!{\langle x_n^2 \rangle} \over {\langle x_{n+l}^2 \rangle}  } (1+2\epsilon)^{l}
\,\,\, \xrightarrow{\!\kappa=0\!} \sqrt{n\over n+l}, 
\eea
which is abbreviated as $ C(n,l|\epsilon,M)$ with the specification of $\epsilon$. 
The corresponding data correlation is measured from the random choices of consecutive 
$x_n$ and $x_{n+l}$ in the data, and it is denoted as $C(n,l) \pm \delta C(n,l)$ (Fig. 4B). 
From the correlation parameter $\kappa$ that is extracted from the histogram, 
we can independently estimate the correlation parameter $\kappa'$ from the auto-correlation. 
For this reason, we also define the discrepancy between the model and the data as 
\bea
E'(\epsilon,n)=\sum_{l=1}^L  \frac{[C(n,l) - C(n,l|\epsilon,M)]^2}{2 \delta^2 C(n,l)}, 
\eea
where the total iteration number is constrained as $N = n+L$. 
Following the previous procedures, we estimate the correlation parameter 
$\kappa'$=0.43 ($n$=5, $L=15$), 0.41 ($n$=10, $L$=10), and 0.38 ($n=15$, $L=5$)
in three cases satisfying the condition, $N=20$.
The measured correlation $\kappa'$ from the auto-correlation is close to the correlation $\kappa$
measured from the histogram.
This approximated overlap between $\kappa$ and $\kappa'$, extracted from
two independent measurements, is remarkable.
%\\
%We have also extracted the correlation ${\hat{C}}_n(L) \pm \delta {\hat{C}}(L)$ of $X_n$ and $X_{n+L}$ in the data. 
%Then we estimate the expectation value of $\epsilon_N$  in the model that could explain the observed correlation.
%Note that $N = n+L$. Again we can define the discrepancy between model and data as a cost:
%\bea
%E_C(\epsilon_N)=\sum_{i=1}^L \Big[ \frac{\hat{C}_n(i) - C_n(i|\epsilon_N)}{\delta \hat{C}_n(i)} \Big]^2. 
%\eea
%Following the above procedures, we estimate $\kappa_C$=0.43 ($n$=5, $L=15$), 0.40 ($n$=10, $L$=10), and 0.40 ($n=15$, $L=5$).
%The estimated correlation strength $\kappa_C$ is similar with the previous estimation $\kappa$ from the histogram fit. 
In Fig. 4B, the auto-correlation of data is matched well with the model for small time lag $l$, 
while the result is matched well with the Bernoulli process for large 
lag large $l$.
In other words, the financial data show strong correlation between events with small time lag.
However, after a certain finite time lag, the correlation becomes negligible so that
the auto-correlation with sufficient time lag converges to the auto-correlation of the Bernoulli process.
To compare which models explain the auto-correlation better within the finite lag window in Fig. 4B,
we compute the Bayes factor between two models of our correlation model $M$ versus the Bernoulli model $B$:
\bea
\frac{P(D|M)}{P(D|B)}=\frac{\int d\epsilon P(D|\epsilon,M) P(\epsilon|M)}{P(D|\epsilon\!=\!0,M)}.
\eea
The Bayes factor, $\ln [P(D|M)/P(D|B)]\!\sim\!10^4$ demonstrates that the correlation model explains
the auto-correlation result exceedingly better than the Bernoulli model in the absence of correlation between events.
 
Our proposed model offers a new scheme of volatility prediction based on the detection of correlations. Unlike ordinary continuous Gaussian volatility prediction models, such as auto-regressive moving average models and generalized auto-regressive conditional heteroskedasticity models, our model can explain finite $N$ statistics, and can predict real data with outstanding precision and  high efficiency due to the analytic result. 

%\section {Acknowledgements}
\acknowledgements
J.K. thanks Jae Sung Lee, Jung-Hyuck Park, Kanghoon Lee, Petre Jizba, Paul Jung, Sang-Woo Lee and Sukjin Yun for discussions, Byoung ki Seo for offering data sets, and Jaeyun Sung for helpful comments to the manuscript. J.J. acknowledges the Max Planck Society, the Korea Ministry of Education, Science and Technology, Gyeongsangbuk-Do and Pohang City for the support of the Junior Research Group 
at APCTP.

\end{document}